\documentclass[12pt]{iopart}

\usepackage{iopams}

\usepackage{epsfig}
\usepackage{sidecap}
\usepackage{color}

\begin{document}
\title[J. Erler et al, ``Misfits in Skyrme-Hartree-Fock'']
{Misfits in Skyrme-Hartree-Fock}

\author{
J. Erler$^{1}$
\footnote[3]{To
whom correspondence should be addressed 
(jochen.erler@physik.uni-erlangen.de)}, P. Kl\"upfel$^1$, P.-G. Reinhard$^{1}$}

\address{$^1$ Institut f\"ur Theoretische Physik,
 Universit\"at Erlangen, D-91058, Erlangen, Germany}

\begin{abstract}
We address very briefly five critical points in the context of
the Skyrme-Hartree-Fock (SHF) scheme: 1) the impossibility to
consider it as an interaction, 2) a possible inconsistency of 
correlation corrections as, e.g., the center-of-mass correction,
3) problems to describe the giant dipole resonance (GDR) 
simultaneously in light and heavy nuclei, 4) deficiencies in
the extrapolation of binding energies to super-heavy elements (SHE),
and 5) a yet inappropriate trend in fission life-times when
going to the heaviest SHE. While the first two points have more a formal
bias, the other three points have practical implications and
wait for solution.
\end{abstract}




\submitto{\JPG}



\section{Introduction}
This manuscript addresses a couple of open problems in the
Skyrme-Hartree-Fock (SHF) scheme. Before coming to these puzzling
details, we want to emphasize the enormous merits of SHF. The SHF
energy functional manages to establish a reliable description of nuclear
properties all over the chart of isotopes (except perhaps the lightest
ones) with an adjustment of only a dozen universal parameters, 
for reviews see e.g. \cite{Ben03aR,Stone_rew}. The enormous
success of SHF implies the temptation to ask for more details and it
is mostly here where we encounter the present limitations of SHF. The
aim of this contribution is to identify problems in order to solve
them later on in a common effort of the nuclear physics community. We
will in the following address five points: the interpretation as
``force'', the consistency of ground state correlations, the giant
dipole resonance in light nuclei, extrapolation of binding energies to
super-heavy elements (SHE), and fission of SHE. The first two points
are of formal nature, the last three more phenomenological.

\section{Formal inconsistencies}

\subsection{The concept of a ``force''}

One often thinks in terms of a ``Skyrme-force'', to be more
precise a Skyrme interaction, whose most touchy 
ingredient is the density dependent term
\begin{equation}
  \hat{V}_3(\mathbf{r}_1,\mathbf{r}_2)
  =
  \frac{t_3}{6}\delta(\mathbf{r}_1-\mathbf{r}_2)
  \rho^\alpha(\mathbf{r}_1)
  \left(1+x_3\hat{\Pi}_{\sigma}\right)
  \quad,\quad
  \rho(\mathbf{r})
  =
  \langle\Phi|\hat\rho(\mathbf{r})|\Phi\rangle
  \quad,
\label{eq:3body}
\end{equation}
where $\hat{\Pi}_{\sigma}$ is the spin-exchange operator.
We argue that this is a most dubious object.  It is not a ``stand
alone'' interaction operator, but depends on a mean-field state
$|\Phi\rangle$ from which the density $\rho(\mathbf{r})$ is taken.
Non-integer values of $\alpha$ immediately hinder an
identification with an $N$-body force.  The simplest case is
$\alpha=1$ and one is tempted to interpret the interaction then as a
three body force.  Let us consider this case and also ignore the
spin-exchange term by setting $x_3=0$ for simplicity. If the operator
(\ref{eq:3body}) was a true interaction operator, one should be able
to produce an equivalent expression in terms of Fermion operators as
\begin{equation}
  \hat{V}_3^\mathrm{(FO)}  
  =
  \sum_{\alpha_1\alpha_2\alpha_3\beta_1\beta_2\beta_3}
  V_{\alpha_1\alpha_2\alpha_3\beta_1\beta_2\beta_3}
  \hat{a}^\dagger_{\alpha_1}
  \hat{a}^\dagger_{\alpha_2}
  \hat{a}^\dagger_{\alpha_3}
  \hat{a}^{\mbox{}}_{\beta_3}
  \hat{a}^{\mbox{}}_{\beta_2}
  \hat{a}^{\mbox{}}_{\beta_1}
  \quad.
\label{eq:V3-FO}
\end{equation}
The ground-state expectation value of such a $\hat{V}_3^\mathrm{(FO)}$
reads
\begin{equation}
  \langle\Phi|\hat{V}_3^\mathrm{(FO)}|\Phi\rangle
  =
  \sum_{nmk}V_{nmk,\widetilde{nmk}}
  \quad,
\label{eq:V3-expFO}
\end{equation}
where $\widetilde{nmk}$ stands for anti-symmetrization of
all three states $nmk$.
However, the expectation value of the effective interaction
(\ref{eq:3body}) for $\alpha=1$ reads
\begin{eqnarray}
  \langle\Phi|\hat{V}_3|\Phi\rangle
  &\propto&
  \sum_{nmk=1}^N
  \int d^3r
  \left[
  \varphi^\dagger_{n}
    \varphi_{n}^{\mbox{}}
  \varphi^\dagger_{m}
    \varphi_{m}^{\mbox{}}
  -
  \varphi^\dagger_{n}
    \varphi_{m}^{\mbox{}}
  \varphi^\dagger_{m}
    \varphi_{n}^{\mbox{}}
  \right]
  \varphi^\dagger_{k}
  \varphi_{k}^{\mbox{}}
\nonumber\\
  &\equiv&
  \sum_{nmk}V_{nmk,\widetilde{nm}\,k}
  \quad.
\label{eq:V3-effexp}
\end{eqnarray}
Note that the state $k$ is not included in the anti-symmetrization.
Thus the whole expression can never be written in the form
(\ref{eq:V3-expFO}) and the Skyrme ansatz (\ref{eq:3body})
cannot be interpreted as an interaction. 

A unique and consistent object is the total energy which turns
out to be a functional of the local density 
$\rho(\mathbf{r})$ and spin density
${\boldsymbol\sigma}(\mathbf{r})$, i.e.
for $\alpha=1$ and $x_3=0$
\begin{eqnarray}
  E_3
  &=&
  \langle\Phi|\hat{V}_3|\Phi\rangle
  =
  \frac{t_3}{24}
  \int d^3r\Big\{
  2\rho^{3} 
  - 
  \rho\left(\rho_n^2+\rho_p^2\right)
  - 
  \rho\left({\boldsymbol\sigma}^2_p+{\boldsymbol\sigma}^2_n\right)
  \Big\}
  \quad.
\label{eq:E3-funct}
\end{eqnarray} 
The main use of the interaction (\ref{eq:3body}) is that 
it serves nicely as a formal generator
for that functional. But any other use is dangerous. Let us consider,
e.g., the residual interaction in RPA. It is deduced from the energy
functional by second functional derivative \cite{Rei-RPA} and reads,
e.g., for pure density variations (no spin excitations)
\begin{equation}
  V^\mathrm{(res)}_3
  =
  \frac{\partial^2E_3}
       {\partial\rho(\mathbf{r_1})\partial\rho(\mathbf{r_2})}
  =
  \frac{t_3}{2}
  \delta(\mathbf{r}_1-\mathbf{r}_2)
  \rho(\mathbf{r}_1) 
  \quad.
\end{equation}
This has a strength factor $t_3/2$ which
is different from the $t_3/6$ of the initial interaction
(\ref{eq:3body}).
Thus the
interaction (\ref{eq:3body}) is not consistently reproducible by
standard many-body techniques.

Most energy-density functionals are plagued by the self-interaction
error \cite{Drei90aB}. It can be checked simply by considering the
case of exactly one particle. Functionals with self interaction then
yield still a non-vanishing energy which is, of course,
unphysical. The functional (\ref{eq:E3-funct}) yields correctly value
zero for the case of one particle and is thus self-interaction
free. That nice feature is achieved by derivation from the interaction
(\ref{eq:3body}). It ought to be mentioned, however, that a derivation
from an interaction is not a necessary condition for constructing
self-interaction free energy functionals.

We thus have seen from two different aspects that the notion of
a Skyrme ``force'' is misleading. The cleanest view of
Skyrme-Hartree-Fock is to derive it from an energy-density functional.
On the other hand, deriving the functional (\ref{eq:E3-funct}) from
the effective interaction (\ref{eq:3body}) avoids the self-interaction
error and provides the spin terms which otherwise would be much
undetermined. It is a matter of phenomenology to check whether the
thus imposed spin terms in the nuclear energy density functional are
supported by phenomenological data.

\subsection{Fragmentation and collective correlations}

The ground state of a nucleus is usually computed with a correction of
the center-of-mass energy. The motivation is that the mean-field state
violates translational invariance and that one needs to consider an
``intrinsic'' state which is obtained by center-of-mass projection.
This projection can be simplified by many-body techniques (second
order Gaussian-Overlap-Approximation \cite{Rei-cm}) to
\begin{equation}
  E_\mathrm{cm} 
  =
  \frac{\langle\Phi|\hat{P}_\mathrm{cm}^2|\Phi\rangle}{2mA} \approx
  30\,\mathrm{MeV}\,A^{-1/3} 
  \quad.
\label{eq:Ecm}
\end{equation}
Both forms (the operator expectation value or the simple estimate) are
widely used and both include the total nucleon number $A$.  Now
consider fusion of two nuclei. Initially, we have a c.m. energy
(\ref{eq:Ecm}) for each nucleus $A_1$ and $A_2$, but finally only one
for the total $A$, i.e.
\begin{equation*}
  E_\mathrm{cm}^\mathrm{(in)}
  =
  E_\mathrm{cm}(A_1)
  +
  E_\mathrm{cm}(A_2)
  \quad\stackrel{?}{\longleftrightarrow}\quad
  E_\mathrm{cm}^\mathrm{(fus)}
  =
  E_\mathrm{cm}(A_1+A_2)
  \quad.
\end{equation*}
That is inconsistent and is particularly puzzling in between
where one does not know which one of the both rules to apply. The problem was already
noted in \cite{Ber80,Ska07} and an interpolation formula was proposed as an ad-hoc remedy.
We want to analyze the case further. Short closer inspection shows that the six initial c.m.  
degrees-of-freedom merge into three final c.m. degrees-of-freedom, two rotational
degrees-of-freedom (for axial symmetry of the final state), and one
quadrupole mode.
\begin{figure}[h!]
\centerline{
\epsfig{figure=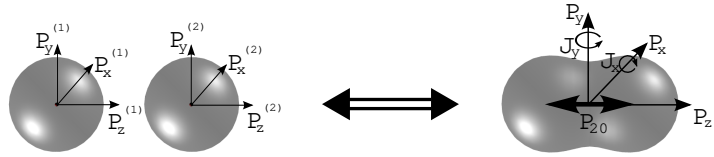,width=0.60\linewidth}
\vspace*{-1cm}
}
\caption{\label{fig:E_cm}
Ambiguity of collective degrees of freedom illustrated for an 
axial symmetric fusion process of two spherical nuclei. 
The initial
six collective c.m. modes ($\{P^{(i)}_x,P^{(i)}_y,P^{(i)}_z\}_{i=1,2}$) 
merge into three c.m. modes ($\{P_x,P_y,P_z\}$) of the compound,
two rotational modes ($\{J_x,J_y\}$) and a quadrupole vibration mode 
($\{P_{20}\}$).
}
\end{figure}
The problem could
be resolved by associating (axial) quadrupole and rotational
correlations with the compound nucleus.  However, 
{this imposes} a new problem: 
we should do the same with the initial two nuclei. This, in
turn, provides even more initial degrees-of-freedom (12 instead of 6)
which have to merge into further collective modes of the compound
system.  This loop generates more and more correlating modes and it is
not clear where to stop.  The problem may be bearable as long as one
considers only intact nuclei with fixed particle number. It becomes a
big hindrance in any reaction which changes particle number.  Thus
there is an urgent need to develop a counting of collective
correlations which is robust under fission, fusion and fragmentation.
For the time being, it is the most consistent procedure to assume that
all correlations are already built into the energy-density functional
and to discard any correlation correction,
even the ones for c.m. or rotational motion. That holds particularly for
all TDHF calculations of large amplitude collective motion.

\section{Trend of the GDR with mass number}

Giant resonances are crucial nuclear excitation modes. They can be
described consistently with a given energy functional by using
time-dependent density functional theory, in the nuclear context
called TDHF, and considering the small amplitude limit thereof.
The scheme is called Random-Phase-Approximation (RPA), for
details see \cite{Rei-RPA}. Information from giant resonances
in heavy nuclei has often been used in the calibration of
a Skyrme parameterization, see e.g. \cite{Brack-skms}. A particular
prominent mode is the Giant Dipole Resonance (GDR) which is commonly
believed to be well under control with SHF. However, this holds
only for heavy nuclei. 
\begin{figure}
\centerline{
\epsfig{figure=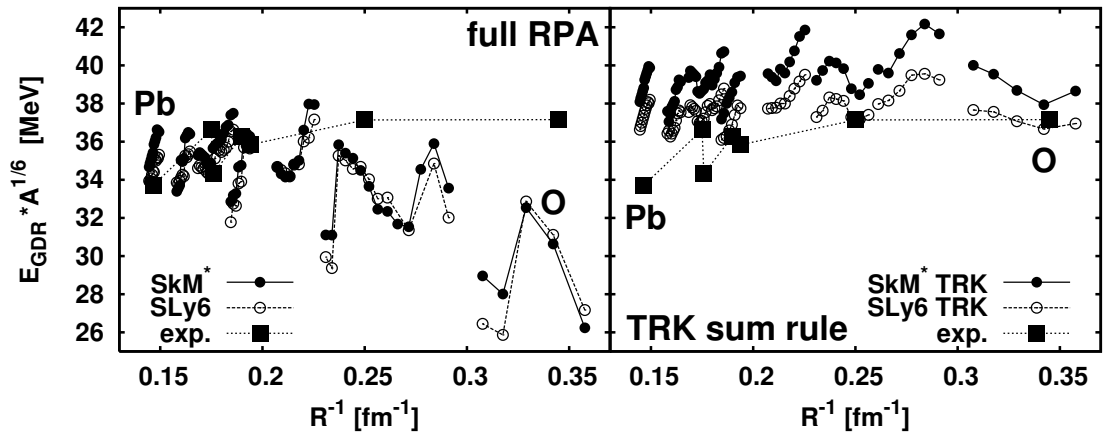,width=0.95\linewidth}
}
\caption{\label{fig:GDR_trends_comp_scale-2}
Peak energies of the Giant-Dipole-Resonance (GDR) drawn versus
inverse radius $R=1.16\,\mathrm{fm}\,A^{1/3}$.
Heavy nuclei(e.g. Pb) are found at the left side and light nuclei
(e.g. O) to the right.
The energies are scaled with $A^{1/6}$. Compared are results
from two different Skyrme parameterizations and experimental
data. 
Left: Peak energy from a full RPA calculation.
Right: Energy from a sum-rule estimate.
}
\end{figure}
This is demonstrated in the left panel of figure
\ref{fig:GDR_trends_comp_scale-2} which compares RPA values for the
average energy of the GDR with experimental data. The peak energies
are deduced from the dipole strength distributions. We show results
for two Skyrme forces (SkM$^*$ \cite{Brack-skms} and SLy6
\cite{sly46}).  We have checked a broad variety of other Skyrme forces
and {always find}
the same trend. The discrepancy is obvious: while the
GDR for heavy nuclei can be adjusted very well, it is impossible to
have simultaneously a reasonable description in small nuclei.
The trend is grossly wrong. The experimental data comply fairly
well with a trend $E_\mathrm{GDR}\propto A^{-1/6}$, but RPA predicts
a much different trend with
a sizeable admixture of $E_\mathrm{GDR}\propto A^{-1/3}$.
The right panel of figure \ref{fig:GDR_trends_comp_scale-2} shows
results of an estimate using the Thomas-Reiche-Kuhn (TRK) sum
rule \cite{Rei-RPA,Brack-GDR}. The peak energy is, of course,
overestimated. But the trend $\propto A^{-1/6}$ complies with
experiment.
The TRK mode is a surface mode (Goldhaber-Teller). The competitor
is the volume mode (Steinwedel-Jensen) which produces a trend
$\propto A^{-1/3}$ \cite{Brack-GDR}. We thus see that the RPA
description underestimates the surface contribution and leaves to much
bias on the volume. The conjecture is that the present Skyrme forces
are still having an inappropriate isovector surface
energy. Substantial improvement in that part is needed.

\section{Extrapolation to SHE}

%
\begin{SCfigure}[0.8][h!]
{
\epsfig{figure=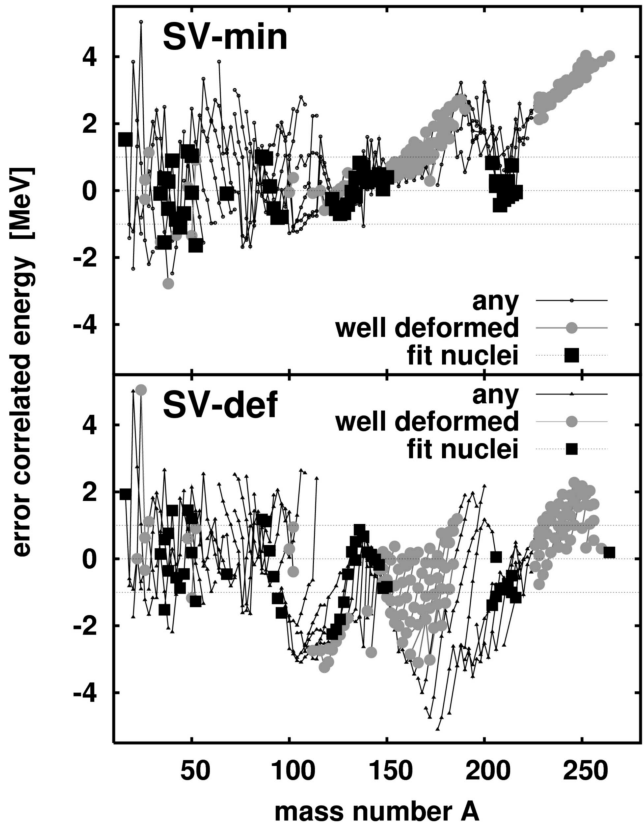,width=9cm}}
\caption{\label{fig:SV-min-def-energies}
Deviation from the experimental binding energies
for a SHF calculation with subsequent correlation corrections
for all available nuclei drawn versus mass number $A$.
The nuclei which were included in the fit of SV-min, or
SV-def respectively, are indicated by filled boxes, 
well deformed nuclei ($\beta_2>0.2$) by grey circles, and
all others by small triangles.
Upper: Results from SV-min.
Lower: Results from SV-def, a force with 
$^{264}$Hs added to the fit data. 
The experimental binding energies were taken from \cite{Aud03}.
}
\end{SCfigure}
Skyrme parameterizations are usually determined by a phenomenological
adjustment to a given pool of fit data (binding energies, radii,
etc., for a chosen set of nuclei). The aim is to obtain a reliable
description for all nuclei deep into the regime of exotic ones. 
The predictive power of a parameterization is to be checked at three
levels:
1) the ability to reproduce the fit-data,
2) the performance for interpolation to other nuclei in the
   range of the fit data, and
3) the reliability of extrapolations to other regions of the
   nuclear chart (e.g. super-heavy elements) or other observables.
It is found that check 1 and check 2 are usually well satisfied while
the extrapolation to super-heavy elements (SHE) reveals a systematic
deviation. That holds for all modern Skyrme parameterizations. We will
discuss {this issue} 
in terms of a newly developed fit protocol
\cite{Klu09a}. Reference point is the Skyrme parameterization SV-min
which was fitted to a large set of nuclei covering a wide span of mass
numbers $A$ as well as long isotopic and isotonic chains. The fit pool
selected good ``mean-field nuclei'', i.e. nuclei which have negligible
effects from collective ground state correlations \cite{Klu08a}, and
was confined to spherical systems for reasons of technical simplicity.

The upper panel of figure \ref{fig:SV-min-def-energies} summarizes the
error in binding energy for SV-min taken over all available nuclei.
All energies are computed including collective ground state
correlations.  Filled squares indicate the fit nuclei (for which
correlations are ignorable), open circles indicate well deformed
nuclei (deformation $\beta_2>0.2$), and open triangles indicate the
majority of vibrationally soft nuclei (vibrational amplitude larger
than deformation, large correlations). The figure shows that
interpolation (results for nuclei $A<210$) works nice with errors
remaining acceptably small and distributed on both sides of the zero
line. But the extrapolation to SHE shows a significant trend to
increasing underbinding.  The same trend (often worse) is found for
other Skyrme forces.

One could try to cure that defect by including data from SHE.  This
has been done by adding the energy of $^{264}$Hs (a well deformed SHE)
to the fit data. This yields a modified parameterization ``SV-def''
whose distribution of errors on binding is shown in the lower panel of
figure \ref{fig:SV-min-def-energies}. The predictions for other SHE
(now being an interpolation) has clearly improved. But that is
achieved at the price of sacrificing the quality of many other
nuclei which are now often overbound. This indicates that
there is an intrinsic problem with the form of the Skyrme energy
functional which inhibits to span a wider mass range.

\section{Fission barriers and half-lives of super-heavy elements}

The microscopic description of nuclear fission is a long standing
problem which was handled long ago in terms of empirical shell models,
see e.g. \cite{Bra72aR}. The case is extremely demanding for
self-consistent mean-field models as all aspects of the effective
nuclear interaction are probed, global parameters of the nuclear
liquid drop as well as details of the shell structure. SHF studies of
fission are thus still rare, see e.g. \cite{Ber01a,Bur04,War06}. We
have recently developed a fully self-consistent description of fission
life-times \cite{Sch09a} and use it here to work out conflicting
trends of fission properties in SHE.  We summarize briefly the
computational scheme as outlined in \cite{Sch09a}:
The fission path is generated by quadrupole-constrained SHF whose
energy expectation values yield a ``raw'' collective energy surface.
The collective mass and moments of inertia are computed by
self-consistent cranking along the states of the path \cite{Rei87aR}.
Approximate projection onto angular momentum zero is performed using
the moments of inertia and angular-momentum width.
Quantum corrections for the spurious vibrational zero-point energy
are applied (using quadrupole mass and width). 
The collective ground state energy is computed fully quantum
mechanically \cite{Klu08a}.
The tunneling rate at the given ground state energy and the repetition
rates are computed by the standard semi-classical formula (known
as WKB) using the quantum-corrected potential energy and collective
mass; the fission life-time is finally composed from these two rates.
All calculations are performed in axial symmetry.
\begin{SCfigure}[0.7]
{
\epsfig{figure=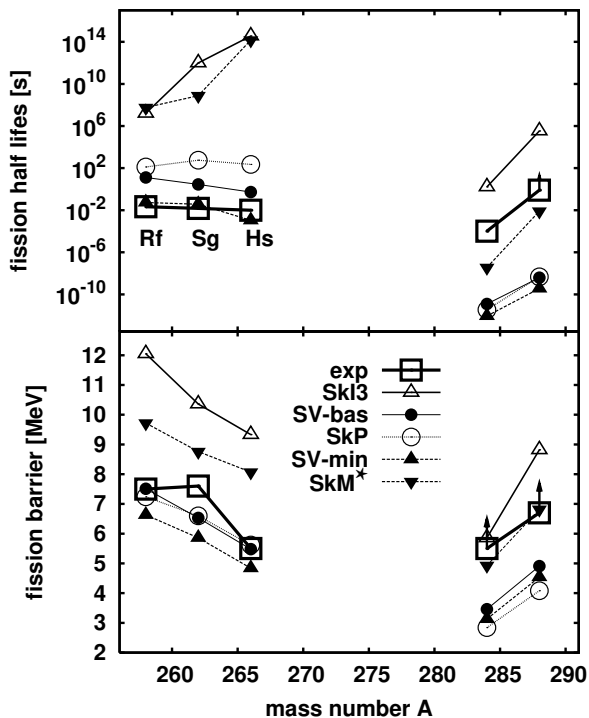,
            width=0.5\linewidth}}
\caption{\label{fig:halflife-barrier3D}
Fission barriers (lower) and half-lives (upper) for a selection
of SHE as indicated. Experimental data are compared with results 
from a variety of different Skyrme
parameterizations, SkM$^*$ \cite{Brack-skms}, SkP \cite{skp},
SkI3 \cite{ski3}, SV-bas and SV-min \cite{Klu09a}.
The experimental data is taken from \cite{Hof01,Pet04,Oga04,Gre06,Gat08}.
} 
\end{SCfigure}
Figure \ref{fig:halflife-barrier3D} summarizes results on fission
barriers and lifetimes for a few typical SHE and for a large variety
of Skyrme parameterizations.  The SHE represent two groups, one at the
lower side and another one with much heavier nuclei at the limits of
present days {available experimental} data.  The span of predictions
from the various Skyrme forces is huge in all cases in spite of the
fact that all these parameterizations provide a high-level description
of basic nuclear properties.  But the variation of predictions is not
the problem.  One may decide to chose from the manifold of
parameterizations just those which provide at the same time good
fission properties throughout. But this turns out to be impossible at
present.  The true problem becomes apparent when looking at the trend
from the lighter side (Rf, Sg, Hs) to the heavier elements (Z=112,
114). All parameterizations produce a wrong trend of the predictions
from the lower to the upper region. 
Forces which perform acceptable for Rf, Sg, Hs fail badly for
Z=112,114 and vices versa.
One may argue that triaxiality,
ignored here, could resolve the trend because triaxial deformation may
lower some barriers selectively.  But that is very unlikely in view of
the experience that the triaxial barrier-lowering amounts typically to
1 MeV, at most 2 MeV \cite{Cwi96a}, which does not suffice to
bridge the gap here.

\section{Conclusion}

We have worked out briefly five puzzling points in connection with
SHF:\\
1) An interpretation as ``force'' is inconsistent because the density
dependence inhibits an expression of the Skyrme energy 
as standard quantum-mechanical expectation value.\\
2) The center-of-mass correction, usually applied, causes conceptual
problems in nuclear fusion, fission and fragmentation;
parameterizations which are used for such reactions should be adjusted
without including the center-of-mass correction.\\
3) It is presently impossible to find a parameterization which
delivers a good description of the giant dipole resonance in all
regions of the nuclear chart; it seems that the relation of surface to
volume mode is not properly balanced.\\
4) The extrapolation of binding energies to SHE yields quickly
increasing underbinding and a refit including energies of known SHE
spoils the quality in the region of stable nuclei; the problem is
probably caused by a still inappropriate surface or curvature energy.\\
5) The experimentally observed trend of fission properties from the Hs
region to much heavier SHE is not reproduced by any SHF
parameterization.\\
For all these points, we do not have presently any solution and often
we have not even figured {out} the deeper reasons. 
This has to be put on the work schedule for future studies.

\bigskip

\noindent
\section*{Acknowledgment}
We thank the regional computing center of the university
Erlangen-N\"urnberg for generous supply of computer time for the
demanding calculations.  The work was supported by the BMBF under
contracts 06 ER 808 and 06 ER 9063.

\end{document}